Terahertz wireless communication at 560-GHz band using Kerr micro-resonator soliton comb

Yu Tokizane,[1,*] Shota Okada,[2] Kenji Nishimoto,[2] Yasuhiro Okamura,[3] Hiroki Kishikawa,[1] Eiji Hase,[1] Jun-Ichi Fujikata,[1] Masanobu Haraguchi,[1] Atsushi Kanno,[4,5] Shintaro Hisatake,[6] Naoya Kuse,[1,7] and Takeshi Yasui[1,*]

[1]Institute of Post-LED Photonics (pLED), Tokushima University, 2-1, Minami-Josanjima, Tokushima 770-8506, Japan

[2]Graduate School of Sciences and Technology for Innovation, Tokushima University, 2-1, Minami-Josanjima, Tokushima 770-8506, Japan

[3]Graduate School of Technology, Industrial and Social Sciences, Tokushima University, 2-1, Minami-Josanjima, Tokushima 770-8506, Japan

[4]National Institute of Information and Communications Technology, 4-2-1 Nukuikitamachi, Koganei, Tokyo 184-8795, Japan

[5]Department of Electrical and Mechanical Engineering, Nagoya Institute of Technology, Gokiso-cho, Showa-ku, Nagoya, Aichi 466-8555, Japan

[6]Electrical and Energy System Engineering Division, Gifu University, 1-1 Yanagido, Gifu 501-1193, Japan

[7]PRESTO, Japan Science and Technology Agency, 4-1-8 Honcho, Kawaguchi, Saitama, 332-0012, Japan

*Corresponding authors: tokizane@tokushima-u.ac.jp, yasui.takeshi@tokushima-u.ac.jp

**Abstract**

Terahertz (THz) waves have attracted attention as carrier waves for next-generation wireless communications (6G). Electronic THz emitters are widely used in current mobile communications; however, they may face technical limitations in 6G with upper-frequency limits. We demonstrate wireless communication in a 560-GHz band by using a photonic THz emitter based on photomixing of a 560-GHz-spacing soliton microcomb in a uni-travelling carrier photodiode together with a THz receiver of Schottky barrier diode. The on-off keying data transfer with 2-Gbit/s achieves a Q-factor of 3.4, thus, satisfying the limit of forward error correction.

Terahertz (THz) waves are a promising candidate for carrier waves in next-generation wireless communications (beyond 5G or 6G, expected carrier frequency > 300 GHz), providing a higher data rate than present wireless communications (5G, carrier frequency = 28 GHz or more) owing to the use of high-frequency carrier waves [1]. In particular, the use of specific THz waves, corresponding to multiple frequency windows available for broadband data transfer in frequencies higher than 350 GHz, is expected for wireless mobile fronthaul and backhaul in 6G for a much higher data rate while avoiding interference with general 6G mobile communication and other applications. One key feature of 6G is its THz emitter. Although electronic THz emitters based on frequency multiplication have been used up to 5G, they may face technical limitations in 6G owing to too high frequency for electronics. For example, higher-order frequency multiplication would largely increase their phase noise owing to the quadratic-dependent increase in phase noise on the number of frequency multiplications. In addition, signal attenuation in electric transmission lines largely increases in such high frequency. These limitations are inherent to electronic THz emitters and may require a different approach.

Another approach for THz emitters is the use of photonic technology because the THz frequency band is located at the boundary between the photonic and electronic regions. Photomixing for optical-to-electric conversion is a promising approach to generate millimeter wave and even THz wave. A pair of near-infrared single-mode CW lasers with an optical frequency separation of THz order interferes with the generation of optical beat signals, and the resulting optical beat signal is then converted into a THz wave by photomixing in a uni-travelling carrier photodiode (UTC-PD) [2,3]. While this method provides higher-frequency THz wave up to a few THz [4], it has a high affinity to optical communication due to fiber-based technology at a telecommunication band of 1.55 μm [5]. In addition, this high affinity enables the easy modulation in the optical region by mature optical modulators and modulation protocols. However, when two independent free-running CW lasers are used for photomixing, fluctuations of frequency and phase in the generated THz wave are synthesized from those of two independent light sources. Such fluctuations often require estimation and/or correction of frequency and phase with a relatively large amount of computation in digital signal processing of THz receiver.

The use of an optical frequency comb (OFC) instead of two CW lasers is a promising approach for the generation of low-phase-nose THz waves, in which two OFC modes with THz frequency spacing are extracted and used for photomixing. Because the frequency spacing $f_{rep}$ of the OFC is inherently stable and its relative phase noise is considerably low owing to mode-locking oscillation and active control in the OFC, the resulting THz wave benefits from the excellent frequency and phase stability in $f_{rep}$. Mode-locked fiber laser (ML-FL) OFCs [6,7] and electro-to-optic-modulator (EOM) OFCs [8-10] have been successfully used for the generation of microwave and millimeter waves. However, when this approach is extended to THz waves, two OFC modes with a large frequency separation (= $mf_{rep}$) must be used for photomixing because the THz wave frequency $f_{THz}$ is much larger than the $f_{rep}$ of these OFCs. For example, when the ML-FL OFC with $f_{rep}$ of 100 MHz is used for generation of 560-GHz THz wave via photomixing, $m$ is achieved to

5600. Such high-order optical frequency multiplication of $f_{rep}$ spoils the low phase noise of $f_{rep}$ characteristic in ML-FL OFC and largely increases the phase noise of the resulting THz wave, similar to the electronic frequency multiplication in electronic THz emitters. In addition, existing OFCs are still bulky, complicated, and expensive for applications in wireless communications.

Recently, the on-chip Kerr micro-resonator soliton comb, namely the soliton microcomb, has attracted attention as a small, simple, and cost-effective OFC benefiting from batch mass production of semiconductor processes. Moreover, micro-resonators largely increase $f_{rep}$ up to a few tens of GHz to a few THz, which is much larger than $f_{rep}$ in ML-FL OFCs or EOM OFCs, while achieving soliton mode-locking oscillation with low phase noise. Such a soliton microcomb enables photomixing of two adjacent OFC modes without the need for optical frequency multiplication for the generation of microwaves [11], millimeter waves [12,13], and even THz waves [14]. Furthermore, a combination of a stabilized microcomb with photomixing generates an ultralow-phase-noise THz wave at 300 GHz [15] and 560 GHz [16]. However, there have been no attempts to apply such microcomb-based THz waves to high-frequency THz wireless communication for wireless mobile fronthaul and backhaul in 6G. In this study, we demonstrate THz wireless communication of on-off keying (OOK) in a 560-GHz band. The 560-GHz THz wave was generated by photomixing of a 560-GHz-spacing soliton microcomb in a UTC-PD, whereas the transmitted THz wave was detected by a Schottky barrier diode (SBD).

We first evaluated the basic performance of the soliton microcomb and the resulting THz wave. Figure 1 shows a schematic of the experimental setup for the generation and evaluation of them. The output light of a single-mode CW laser (pump laser, TLX-1, Thorlabs, Inc., wavelength = 1560.7 nm, optical frequency = 192.09 THz, output power = 20 mW) after passing through a dual-parallel Mach-Zehnder modulator (DP-MZM, FTM7962EP, Fujitsu Optical Components Lim., data rate = 40 Gbps) was amplified by a homemade erbium-doped fiber amplifier (EDFA, output power = 500 mW). The resulting pump light was coupled into a $Si_3N_4$ (SiN) micro-resonator (custom, LIGENTEC, S.A., free spectral range = 560 GHz). To generate a single soliton microcomb with an $f_{rep}$ of 560 GHz, the DP-MZM rapidly tunes the optical frequency of the pump light around the resonant frequency of the micro-resonator [17]. We rejected the residual pump light from the generated soliton microcomb by a homemade optical notch filter (ONF, stop wavelength = 1560.7 nm, stop optical frequency = 192.09 THz). The resulting soliton microcomb (total power = 1 mW) was further amplified to 100 mW using a commercialized EDFA (EDFA100P, Thorlabs, Inc., output power > 20 dBm). Parts of the soliton microcomb were fed into an optical spectrum analyzer (OSA, AQ6370D, Yokogawa Electric Corp., minimum wavelength = 600-1700 nm) for optical spectrum measurement and a two-wavelength delayed self-heterodyne interferometer (TWDI) for the phase noise evaluation of its $f_{rep}$ [18-20]. The residual of the soliton microcomb was incident onto an antenna-integrated UTC-PD (IOD-PMAN-13001-2, NTT Electronics Corp., output frequency of 200–600 GHz, and output power of 10 μW at 600 GHz) via optical fiber to generate THz waves by photomixing. The generated THz wave passed through a pair of Teflon lenses (LAT100, Thorlabs, Inc., focal length = 91 nm) and was incident on a sub-harmonic mixer (SHM, WR1.5SHM, Virginia Diodes, Inc., LO frequency = 250–375 GHz, RF frequency = 500–750 GHz, IF frequency = DC–40 GHz). SHM generates a heterodyned beat signal (frequency = $f_{IF}$) between THz waves (frequency = $f_{THz} = f_{rep}$) and the 54-th higher harmonic component (frequency = $54f_{LO}$) of a local oscillator signal synthesized by a microwave frequency synthesizer (E8257D, Agilent Tech., Inc., frequency = $f_{LO}$ = 10.37 GHz). The resulting RF beat signal was measured using an RF spectrum analyzer (RSA, 89441A, Agilent Tech., Inc., frequency = DC–2.65 GHz) for the spectrum observation of $f_{THz}$.

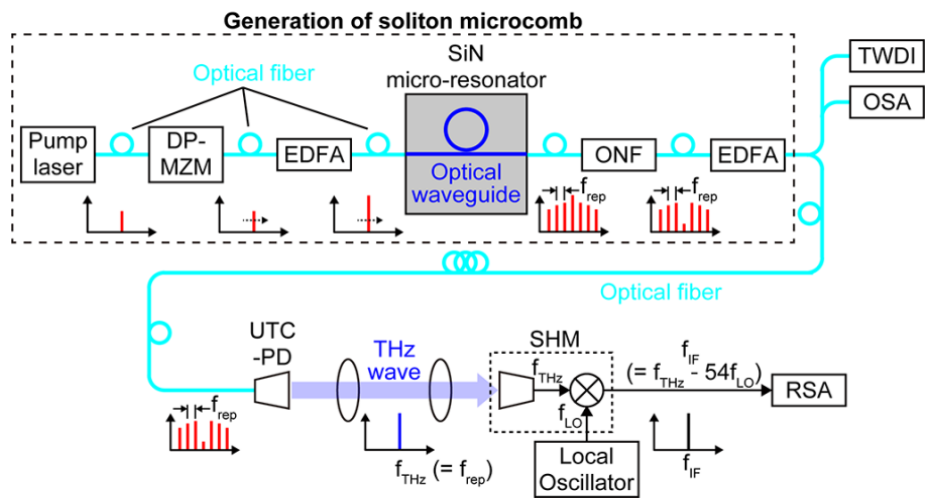

Fig. 1. Experimental setup for generation and evaluation of a soliton microcomb and THz wave.

Figure 2(a) shows the optical spectrum of the soliton microcomb after passing through ONF (resolution bandwidth = 0.2 nm). We confirmed that multiple OFC modes were distributed at intervals of 560 GHz, around the wavelength of the pump light. In addition, a broad spectrum of amplified spontaneous emission (ASE) appears as the optical background signal of the soliton microcomb, which influences the optical signal-to-noise ratio (OSNR) of each OFC mode. We later discuss the influence of such an ASE background on the quality of data transfer in OOK wireless communications. Figure 2(b) shows the single sideband (SSB) phase noise power spectrum density (PSD) of $f_{rep}$, corresponding to the relative phase noise between the two OFC modes, measured by TWDI. The SSB phase noise of our soliton microcomb is -60 dBc/Hz at an offset frequency of 10 kHz, which can be completely transferred to the THz wave generated from the soliton microcomb without additional phase noise in UTC-PD [16]. To determine $f_{THz}$ and investigate its spectral features in detail, we measured the RF spectrum of the heterodyned beat signal (frequency = $f_{IF}$) between $f_{THz}$ and 54$f_{Lo}$ (resolution bandwidth or RBW = 1 MHz), as shown in Fig. 2(c). The resulting RF spectrum indicated a spectral peak at 150 MHz (=$f_{IF}$). We determined $f_{TH}$ to be 560.13 GHz from the measured $f_{IF}$ and the known $f_{LO}$; it is exactly equal to $f_{rep}$.

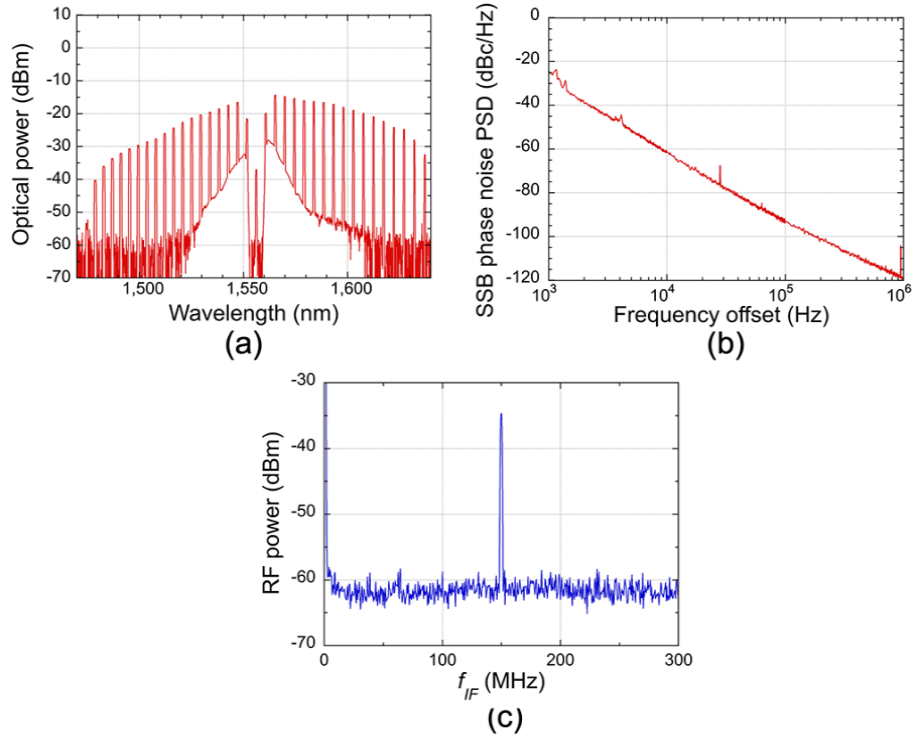

Fig. 2. Fundamental property of a soliton microcomb and THz wave. (a) Optical spectrum and (b) phase noise of $f_{rep}$ in a 560-GHz-spacing microcomb. (c) RF spectrum of heterodyned beat signal.

We next demonstrated OOK wireless communication with a 560-GHz THz wave. Figure 3 shows a schematic of the experimental setup for the THz wireless communication. We generated a soliton microcomb without residual pump light ($f_{rep}$ = 560 GHz, optical power = 100 mW) using the experimental setup of microcomb generation in Fig. 1. Two adjacent microcomb modes with a relatively higher OSNR (OFC-M1, wavelength = 1533.87 nm, optical frequency = 195.448 THz, optical power = 3.7 mW; OFC-M2, wavelength = 1538.28 nm, optical frequency = 194.888 THz, optical power = 2.1 mW) were extracted from the soliton microcomb using a programmable optical filter

(WaveShaper 4000S, Finisar, filter bandwidth = 10 GHz). One of the two extracted microcomb modes (OFC-M1) was delivered to a LiNbO$_3$ modulator (LN-MOD, T.MXH1.5-20PD-ADC-LV, Sumitomo Osaka Cement Co., Ltd, wavelength = 1.55 μm, modulation rate = 40 Gbit/s, optical bandwidth > 20 GHz) for OOK modulation. The OOK signal of a 2-Gbit/s non-return-to-zero (NRZ) signal with a DC bias, generated by an arbitrary waveform generator (AWG, M8196A, Keysight Tech., Inc., sample rates < 92 GS/s, analog bandwidth = 32 GHz), was applied to the LN-MOD as a modulation signal. The OOK-modulated OFC-M1 was amplified by an EDFA, and the residual ASE background outside was removed by a tunable optical bandpass filter (OBPF, WTF-200, Alnear Inc., optical bandwidth < 0.3 nm). The other microcomb mode (OFC-M2) passed through another EDFA and OBPF without OOK modulation. The modulated OFC-M1 and unmodulated OFC-M2 were combined by a fiber coupler, and the resulting two microcomb modes (total power = 30 mW) were fed into the UTC-PD to generate a THz wave at an $f_{THz}$ of 560 GHz. The OOK-modulated THz wave propagated in free space (optical pathlength = 0.6 m) via a pair of Teflon lenses and was then detected by SBD (Virginia Diodes, Inc., WR1.9ZBD, RF frequency = 400–600 GHz, responsivity = 1000 V/W). The electric output signal from the SBD was amplified using an electric amplifier (AMP, ZX60-V62+, Mini-Circuits, frequency range = 0.05–6 GHz, gain = 14.4 dB at 6 GHz) via a bias-T (not shown), and was then measured by a real-time oscilloscope (UXR0402AP, Keysight Tech., Inc., max. bandwidth = 40 GHz, max. sample rate = 256 GSa/s).

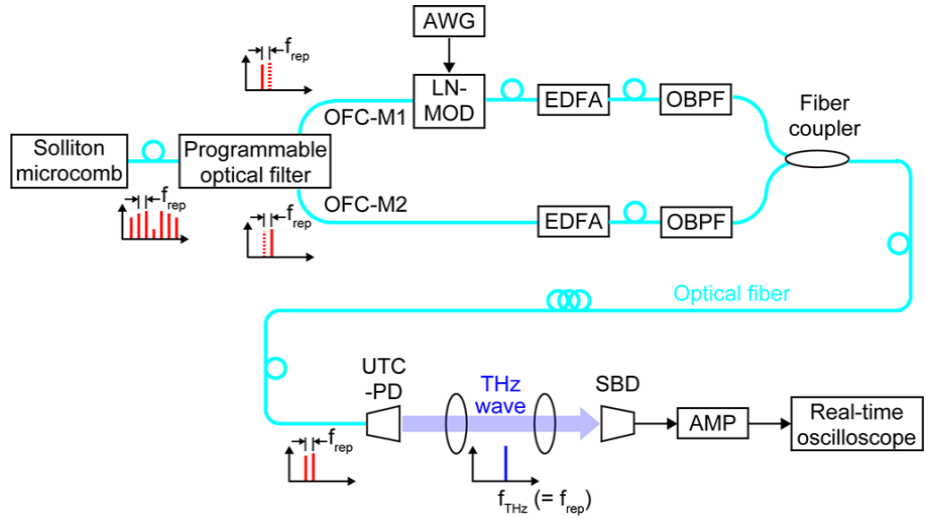

Fig. 3. Experimental setup for OOK wireless communication at 560 GHz.

To suppress the ASE background while securing the required optical amplification of microcomb modes, an optical input signal with a suitable power must be fed into EDFAs because an extremely weak optical input increases ASE unnecessarily. We precisely adjusted the optical input signal of the EDFAs for the modulated OFC-M1 and unmodulated OFC-M2 for a better OSNR. We further reduced the ASE background outside of two OFC modes by fine tuning of OPBFs. Figure 4(a) shows the optical spectrum of the two microcomb modes immediately before being sent to the UTC-PD (RBW = 0.2 nm). These microcomb modes appear to have a sufficient dynamic range (= 40 dB) from the ASE background level around -30 dBm. For comparison, instead of separately amplifying and filtering with EDFAs and OPBFs before fiber coupling, we simultaneously amplified the two microcomb modes by a single EDFA after their fiber-coupling. In this case, the broad ASE background remains outside of two microcomb modes even though OSNR was achieved to 40 dB, as shown in Fig. 4(b). Such difference of ASE background between them in the same OSNR influences the result of OOK wireless communication described later.

The 560-GHz THz wave, generated by photomixing of two microcomb modes in Fig. 4(a), was used for OOK wireless communication. Figure 4(c) shows an eye diagram of OOK with 2 Gbit/s measured by the real-time oscilloscope; thus, the eye was opened. The Q-factor of this eye pattern was determined to be 3.40. This Q-factor satisfied the limit of forward error correction (FEC limit, Q-factor = 2.33). Accordingly, 2-Gbit/s OOK THz wireless communication was successfully demonstrated in the 560-GHz band within the FEC limit. For comparison, if two microcomb modes in Fig. 4(b) were used for THz generation, the eye was fully closed (not shown). In this way, two pairs of EDFAs and OBPFs separately for two microcomb modes plays an important role in the present wireless communication.

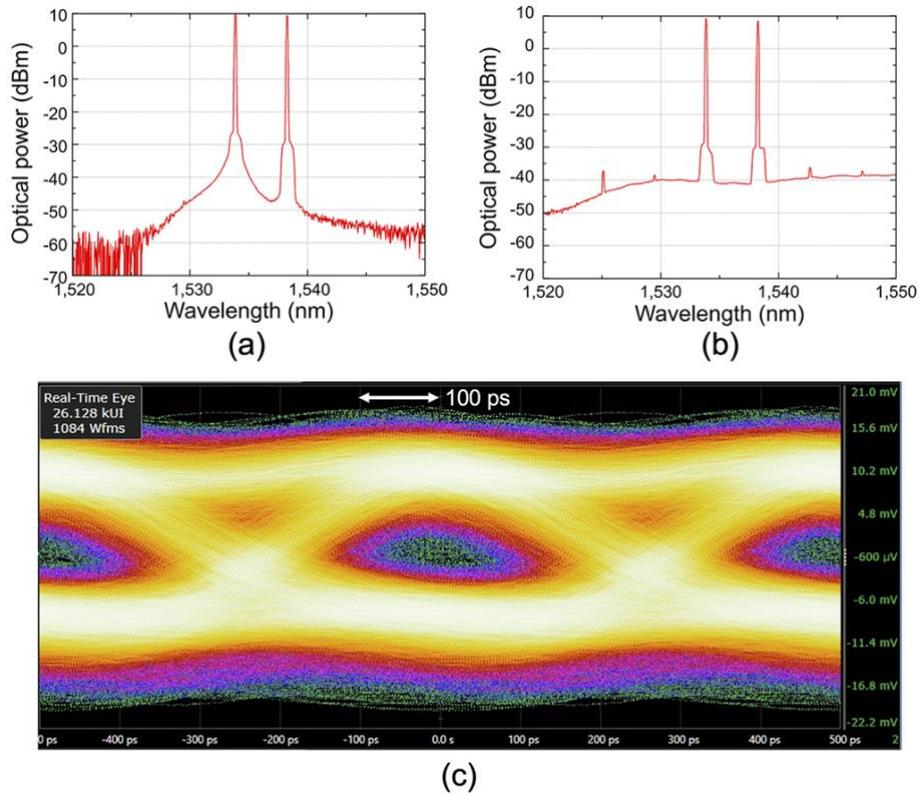

Fig. 4.   Results of 2-Gbit/s OOK wireless communication at 560 GHz.   Optical spectrum of two microcomb modes obtained by (a) two pairs of EDFAs and OPBFs before fiber coupling and (b) a single EDFA after their fiber-coupling. (c) Eye pattern of OOK signal with 2 Gbit/s.

We discuss the causes of the relatively low Q-factor of the measured OOK signal. One possible reason for it is atmospheric attenuation caused by water vapor absorption. The 560-GHz band is close to the water absorption line of $1_{10}$-$1_{01}$ (center frequency = 557 GHz, pressure broadening linewidth ∼ several GHz at atmospheric pressure). As the attenuation coefficient of water vapor is 7.1 dB/m at 560 GHz [21], the 0.6-m propagation of THz wave causes the THz attenuation by 4.26 dB. Use of such considerably attenuated

THz wave limits the present Q factor. If THz wave at 565 GHz is used in place of 560 GHz, the atmospheric attenuation reduces to 1.14 dB from an attenuation coefficient of 1.9 dB/m [21]. In this case, OOK wireless communication will largely enhance its Q-factor. Fortunately, in the photomixing of a soliton microcomb with UTC-PD, it is easy to change $f_{rep}$, and hence $f_{THz}$ because multiple micro-resonators with different $f_{rep}$ can be integrated in a single chip.

In conclusion, we demonstrated 2-Gbit/s OOK wireless communication in the 560-GHz band based on a hybrid system of photomixing-based photonic THz generation and SBD-based electric THz detection. Photomixing of two adjacent microcomb modes via UTC-PD enables the generation of THz waves without the need for optical frequency multiplication. Although such a higher-frequency band is accessible by UTC-PD-based photomixing, the combination of photomixing with the soliton microcomb benefits from the inherently low phase noise of $f_{rep}$ in the soliton microcomb. A pair of OOK-modulated and unmodulated microcomb modes with an OSNR over 40 dB was used to generate the OOK-modulated THz wave. The achieved Q-factor was 3.40, which was within the range of the FEC limit.

The achieved phase noise of $f_{rep}$ remained around -60 dBc/Hz at an offset frequency of 10 kHz because of the use of the free-running soliton microcomb; however, there is still room to further decrease the phase noise down to -99 dBc/Hz at an offset frequency of 10 kHz by active control of the soliton microcomb [16]. The introduction of such ultra-low phase noise soliton microcombs in the current wireless communication system will help in achieving not only error-free OOK data transfer without the help of FEC but also advanced-modulation-format data transfers, such as PSK, QAM, and QPSK, owing to the ultralow phase noise of the THz wave.

**References**

1. S. Dang, O. Amin, B. Shihada, and M.-S. Alouini, " What should 6G be?," Nat. Electron. **3**, 20 (2020).

2. H. Ito, T. Furuta, S. Kodam, and T. Ishibashi, " InP/InGaAs uni-travelling-carrier photodiode with 310 GHz bandwidth," Electron. Lett. **36**, 1809 (2000).

3. T. Ishibashi and H. Ito, "Uni-traveling-carrier photodiodes ," J. Appl. Phys. **127**, 031101 (2020).

4. D. Fukuoka, K. Muro, and K. Noda, "Coherent THz light source based on photo-mixing with a UTC-PD and ASE-free tunable diode laser," Proc. SPIE **9747**, 974717 (2016).

5. H. Song, K. Ajito, Y. Muramoto, A. Wakatsuki, T. Nagatsuma, and N. Kukutsu, "24 Gbit/s data transmission in 300 GHz band for future terahertz communications," Electron. Lett. **48**, 19 (2012).

6. S. A. Diddams, A. Bartels, T. M. Ramond, C. W. Oates, S. Bize, E. A. Curtis, J. C. Bergquist, and L. Hollberg, "Design and control of femtosecond lasers for optical clocks and the synthesis of low-noise optical and microwave signals," IEEE J. Sel. Top. Quantum Electron. **9**, 1072 (2003).

7. T. M. Fortier, M. S. Kirchner, F. Quinlan, J. Taylor, J. C. Bergquist, T. Rosenband, N. Lemke, A. Ludlow, Y. Jiang, C. W. Oates, and S. A. Diddams, "Generation of ultrastable microwaves via optical frequency division," Nat. Photon. **5**, 425 (2011).

8. S. Xiao, L. Hollberg, and S. A. Diddams, "Low-noise synthesis of microwave and millimetre-wave signals with optical frequency comb generator," Electron Lett. **45**, 170 (2009).

9. G. Qi, J. Yao, J. Seregelyi, S. Paquet, and C. Bélisle, "Generation and distribution of a wide-band continuously tunable millimeter-wave signal with an optical external modulation technique," IEEE Trans. Microw. Theory Tech. **53**, 3090 (2005).

10. A. Ishizawa, T. Nishikawa, T. Goto, K. Hitachi, T. Sogawa, and H. Gotoh, "Ultralow-phase-noise millimetre-wave signal generator assisted with an electro-optics-modulator-based optical frequency comb," Sci. Rep. **6**, 24621 (2016).

11. W. Liang, D. Eliyahu, V. S. Ilchenko, A. A. Savchenkov, D. Seidel, L. Maleki, and A. B. Matsko, "High spectral purity Kerr frequency comb radio frequency photonic oscillator," Nat. Commun. **6**, 1 (2015).

12. J. Liu, E. Lucas, A. S. Raja, J. He, J. Riemensberger, R. N. Wang, M. Karpov, H. Guo, R. Bouchand, and T. J. Kippenberg, "Photonic microwave generation in the K-band using integrated soliton microcombs," Nat. Photon. **14**, 486 (2020).

13. J. Li, H. Lee, T. Chen, and K. J. Vahala, "Low-pump-power, low-phase-noise, and microwave to millimeter-wave repetition rate operation in microcombs," Phys. Rev. Lett. **109**, 233901 (2012).

14. S. Zhang, J. M. Silver, X. Shang, L. D. Bino, N. M. Ridler, and P. Del'Haye, "Terahertz wave generation using a soliton microcomb," Opt. Express **27**, 35257 (2019).

15. T. Tetsumoto, T. Nagatsuma, M. E. Fermann, G. Navickaite, M. Geiselmann, and A. Rolland, "Optically referenced 300 GHz millimetre-wave oscillator," Nat. Photon. **15**, 516–522 (2021).

16. N. Kuse, K. Nishimoto, Y. Tokizane, S. Okada, G. Navickaite, M. Geiselmann, K. Minoshima, and T. Yasui, "Low phase noise THz generation from a fiber-referenced Kerr microresonator soliton comb," Commun. Phys., to be accepted; arXiv:2204.02564 (2022).

17. K. Nishimoto, K. Minoshima, T. Yasui, and N. Kuse, "Investigation of the phase noise of a microresonator soliton comb," Opt. Express **28**, 1929 (2020).

18. K. Jung and J. Kim, "All-fibre photonic signal generator for attosecond timing and ultralow-noise microwave," Sci. Rep. **5**, 16250 (2015).

19. N. Kuse and M. E. Fermann, "Electro-optic comb based real time ultra-high sensitivity phase noise measurement system for high frequency microwaves," Sci. Rep. **7**, 2847 (2017).

20. N. Kuse and M. E. Fermann, "A photonic frequency discriminator based on a two wavelength delayed self-heterodyne interferometer for low phase noise tunable micro/mm wave synthesis," Sci. Rep. **8**, 13719 (2018).

21. P. Baron, J. Mendrok, Y. Kasai, S. Ochiai, T. Seta, K. Sagi, K. Suzuki, H. Sagawa, and J. Urban, "AMATERASU: model for atmospheric terahertz radiation analysis and simulation," J. National Inst. Info. Commun. Tech. **55**, 109 (2008).